# Bounds on the Sum Capacity of Synchronous Binary CDMA Channels

K. Alishahi[1], F. Marvasti[2], V. Aref[2], P. Pad[2]

**Abstract**

In this paper, we obtain a family of lower bounds for the sum capacity of Code Division Multiple Access (CDMA) channels assuming binary inputs and binary signature codes in the presence of additive noise with an arbitrary distribution. The envelope of this family gives a relatively tight lower bound in terms of the number of users, spreading gain and the noise distribution. The derivation methods for the noiseless and the noisy channels are different but when the noise variance goes to zero, the noisy channel bound approaches the noiseless case. The behavior of the lower bound shows that for small noise power, the number of users can be much more than the spreading gain without any significant loss of information (overloaded CDMA). A conjectured upper bound is also derived under the usual assumption that the users send out equally likely binary bits in the presence of additive noise with an arbitrary distribution. As the noise level increases, and/or, the ratio of the number of users and the spreading gain increases, the conjectured upper bound approaches the lower bound. We have also derived asymptotic limits of our bounds that can be compared to a formula that Tanaka obtained using techniques from statistical physics; his bound is close to that of our conjectured upper bound for large scale systems.

---

[1] Department of Mathematical Sciences, Sharif University of Technology, Tehran, Iran. kasraalishahi@math.sharif.edu.

[2] Advanced Communications Research Institute (ACRI), Sharif University of Technology, Tehran, Iran. marvasti@sharif.edu.

# 1. Introduction

Multiple Access Channels (MAC) with many users and Multi-User Detection (MUD) at the receiver give rise to information theoretic problems and concepts much more complicated than the classical situation of single user channels. Although comprehensive theorems for capacity regions have been developed for MAC, an explicit computation of the capacity region is not known in terms of specific model parameters. In this paper, we intend to derive a relatively tight family of lower bounds and a conjectured upper bound for a binary Multi-User CDMA with binary signatures; we assume synchronous CDMA and additive Independent and Identically Distributed (*i.i.d.*) noise with any distribution. Below we will give a brief historical development of this area:

The capacity region consists of a set of information rates such that simultaneous reliable communication is possible for each user. This problem was developed by Ahlswede [l]-[2] and Liao [3] on the two-user discrete memoryless channel. An explicit expression for the capacity region of the Gaussian discrete memoryless MAC was given by Cover [4] and Wyner [5] and discussed in [6].

Verdu in [7] found the capacity region of the CDMA channel as a function of the cross-correlations between the assigned signature waveforms and their signal-to-noise ratios for the symbol synchronous case and for inputs with power constraints. The same author [8] found the capacity region for symbol asynchronous case for Gaussian distributed inputs with power constraints; in these two papers, Verdu showed that the achievable rates depend only on the correlation matrix of the spreading coefficients. These issues and complexity of MUD receivers were discussed by the same author in his book [9]. In [10], the authors considered random spreading and analyzed the spectral efficiency, (defined as bits per chip that can be transmitted reliably) for linear detectors. In the limit, when the number of users and the signature length go to infinity (the ratio is kept constant), they obtained nice analytical formulas for the spectral efficiency and showed from concentration theorems that the spectral efficiency of a typical random selection of signature matrices and its average are the same with high probability. These formulas follow from the known spectrum of large covariance matrices.

For finite number of users with real-valued inputs and real valued signatures, an upper bound for the sum capacity has been defined in [18]. The extension of the sum capacity bounds for asymmetric user power constraint is given in [19]; they have also identified the sequences that achieve the sum capacity. The authors of reference [20] have found upper and lower bounds for the spectral efficiency (defined as the sum capacity by the authors) under quasi-static fading environments, channel estimation, and training sequences; the bound evaluations are based on the works of [21]-[22].

For the binary input values, not much is known except the asymptotic behavior for the spectral efficiency in the limiting case where the number of users ($n$) and the spreading gain ($m$) go to infinity when the ratio $n/m$ is kept constant and SNR values are large [11]. The random matrix techniques used for Gaussian inputs do not apply here because the spectral efficiency cannot be written in terms of just the covariance matrix of the spreading sequences. Tanaka [12] applied a technique from statistical mechanics to this problem and derived a formula for the normalized sum capacity. Tanaka evaluated the performance of a class of CDMA multiuser detectors in the large-system limit analytically. These results were later extended in [13] to include unequal powers and fading channels. This method is non-rigorous but later, Montanari and Tse [14] have made progress towards a rigorous derivation of Tanaka's capacity formula.

The authors of reference [15] have shown that, for large systems, the capacity concentrates around its mean, i.e. a random signature matrix results in a capacity very close to the "mean capacity"[3] with high probability. The same authors in [16] claimed that Tanaka's formula is an upper bound to the capacity for all values of the parameters and derived various concentration theorems for the large-system limit.

For binary inputs and random binary CDMA channels, we have derived a relatively tight lower bound for the sum capacity of the noiseless case [17] and [30]. In these references, we have shown that interference free overloaded CDMA is possible. In what follows, we will derive the sum capacity for the noiseless and noisy scenarios. Our closed form derivations, unlike the previous works, do not depend on the limiting cases but rather on the number of users, spreading gain (signature length), and the noise distribution. The derivation of the noisy case is for a general *i.i.d.* distribution and special cases such as Gaussian and uniform distributions are also derived in closed forms.

---
[3] The mean is with respect to the randomness of the signature matrix.

In the next section, the preliminaries and some propositions are discussed. Section 3 is the main part of the paper where the supremum of a family of lower bounds for noisy channels is derived. The sum capacity lower bounds for Gaussian and uniformly distributed additive noise are derived as special cases. Section 4 covers a derivation of a conjectured upper bound for the sum capacity that is close to the lower bounds, and hence the adjective of "relatively tight" bounds. Section 5 gives the asymptotic derivation of the normalized sum-capacity when the number of users ($n$) and the spreading gain ($m$) go to infinity while the ratio $n/m$ is kept constant; a comparison of our results with that of Tanaka is also included. Simulation results are discussed in Section 6, and finally the concluding remarks are in Section 7.

## 2. Preliminaries

### 2.1. Capacity region

In a MAC, there are several users sending information to a common receiver and the users should overcome not only the noise but also their mutual interference. The "capacity region" ($\mathcal{R}$) of a MAC is defined as the closure of all achievable rate vectors[4]. Ahlswede and Liao [1-3] characterized the structure of the capacity region of an *n*-user discrete memoryless MAC as the closure of the convex hull of the rate vectors $(r_1, r_2, \ldots, r_n)$ satisfying

---

[4] A vector $(r_1, r_2, \ldots, r_n)$ is called an achievable rate vector if it is possible for the senders, using appropriate codes, to send information by these rates with arbitrarily small probabilities of error.

$$\sum_{i\in I} r_i \leq I\left(\{X_i\}_{i\in J}\,;\, Y|\{X_j\}_{j\in J^c}\right) \text{ for all } J \subseteq \{1,2,\dots,n\} \tag{1}$$

for some input product distribution $p_1(x_1)p_2(x_2)\dots p_n(x_n)$; where $X_i$ refers to the symbol sent by user $i$ and $Y$ is the output of the channel.

If we are to assign a single value to a MAC as a measure of capacity of the channel, the sum capacity is the most appropriate candidate. The sum capacity measures the maximum of the total information rate (i.e., the sum of all the user rates) that can be achieved, and is equal to $\max I(X_1, X_2, \dots, X_n; Y)$ where the maximum is taken over all product distributions $p_1(x_1)p_2(x_2)\dots p_n(x_n)$ [23].

## 2.2. Code Division Multiple Access (CDMA) channels

We consider a synchronous CDMA channel as a special case of MACs, in which there are $n$ users sending binary symbols to a common receiver. The $i^{th}$ user has a signature sequence $\boldsymbol{a}_i = [a_{1i}, \dots, a_{mi}]^T$ assumed to be known to the receiver. If $X = [X_1, \dots, X_n]^T$ is the vector of user symbols, the $i^{th}$ user sends $X_i \times \boldsymbol{a}_i$, where $X_i = \pm 1$. Moreover, there exists a background noise $N = [N_1, \dots, N_m]^T$ such that $N_i$'s are *i.i.d.* random variables with any arbitrary distribution. We assume that the channel attenuation is normalized to $\frac{1}{\sqrt{m}}$ (perfect near/far attenuation compensation). Under these assumptions and if the received signal is $Y = [Y_1, \dots, Y_m]^T$, then $Y_j = \frac{1}{\sqrt{m}}\sum_i a_{ji} X_i + N_j$ or equivalently $Y = \frac{1}{\sqrt{m}}AX + N$ where $\boldsymbol{A} = [\boldsymbol{a}_1, \dots, \boldsymbol{a}_n]$ is the $m \times n$ signature matrix. In this paper, we are interested in binary CDMA signature matrices, i.e., $a_{ji} = \pm 1$ or alternatively, the signature matrix is chosen from $\mathcal{M}_{m\times n}(\pm 1)$, the set of all $m \times n$ matrices with $\pm 1$ entries.

## 3. Lower Bounds for the Sum Channel Capacity

In this section we will obtain a family of lower bounds for the sum channel capacity of binary CDMA signature matrix in terms of $m$, $n$ and the noise model. The point is that for given values of $m$ and $n$, we still have the choice of designing the signature matrix in order to optimize the capacity. Therefore, the relevant quantity is the maximum taken over all potential signature matrices in $\mathcal{M}_{m\times n}(\pm 1)$.

In section 3.1, we consider the noiseless channel and derive a lower bound, the behavior of which shows that for a given spreading gain $m$, there is a threshold $n_m$ much larger than $m$, such that while the number of users is less than $n_m$, there exist signature matrices with sum capacities close to $n$.

In section 3.2, we extend our techniques to cover binary CDMA channel with additive *i.i.d.* noise. As special cases, we will obtain lower bounds for channels with Gaussian white noise and for *i.i.d.* noise with uniform distribution.

**3.1. Noiseless channels**

Even in the absence of noise, multi-user interference can affect the channel capacity [17] and [30]. In order to evaluate a lower bound, let $m$ and $n$ be natural numbers. For a matrix $A \in \mathcal{M}_{m\times n}(\pm 1)$, denote the sum channel capacity by $C(A)$, i.e., $C(A) = \max_{p(x_1).p(x_2)....p(x_n)} I(X_1, X_2, ..., X_n; Y|A)$. Now define

$$C(m,n) = \max_{A \in \mathcal{M}_{m\times n}(\pm 1)} C(A)$$

**Theorem 1:** For any $m$ and $n$,

$$C(m,n) \geq n - \log\left(\sum_{j=0}^{\lfloor\frac{n}{2}\rfloor} \binom{n}{2j}\left(\frac{\binom{2j}{j}}{2^{2j}}\right)^m\right) \qquad (2)\,[5]$$

**Proof:** $C(A)$ is the maximum mutual information over all product distributions on $X$. Specifically, $C(A) \geq I(X; Y)$, where $Y = \frac{1}{\sqrt{m}}AX$ and $X$ has uniform distribution on $\{\pm 1\}^n = \{(x_1, ..., x_n)^T : x_i = \pm 1\}$.

---

[5] Throughout this paper, the unit of capacity is in bits and the base of log is 2.

But since the channel is noiseless, the mutual information is simply equal to $H(Y)$ (deterministic channel). In the remaining of the proof, we will find a lower bound for $H(Y)$.

For a given $A$, let $\varphi(x) = \frac{1}{\sqrt{m}} Ax$ be the map induced by $A$, and suppose that $|\varphi(\{\pm 1\}^n)| = k$ and the preimages of the $k$ values of $\varphi(\{\pm 1\}^n)$ have cardinalities $n_1, n_2, \ldots, n_k$. It can be easily seen that the mutual information is equal to

$$H(Y) = -\sum_j \frac{n_j}{2^n} \log \frac{n_j}{2^n}$$

The key point is that the above expression can be rewritten in another form in terms of the input values rather than the output distribution:

$$H(Y) = \frac{-1}{2^n} \sum_{x \in \{\pm 1\}^n} \log \frac{n_x}{2^n}$$

where $n_x = |\varphi^{-1}(\varphi(x))|$ is the number of points mapped by $\varphi$ to the same value as $\varphi(x)$.

Now if we let $A$ to be random with independent entries that are uniformly chosen to be $\pm 1$ and take the expectation of $H(Y)$, we obtain

$$\mathbb{E}_A(H(Y)) = \frac{-1}{2^n} \sum_{x \in \{\pm 1\}^n} \mathbb{E}_A \left( \log \frac{n_x}{2^n} \right)$$

But since there is a symmetry between the elements of $\{\pm 1\}^n$, all terms in the above summation are equal and hence for any $x \in \{\pm 1\}^n$,

$$\mathbb{E}_A(H(Y)) = -\mathbb{E}_A \left( \log \frac{n_x}{2^n} \right) = n - \mathbb{E}_A(\log n_x)$$

Since $\log(z)$ is a concave function of $z$, according to the Jensen's inequality, $\mathbb{E}(\log n_x) \leq \log(\mathbb{E}(n_x))$. Therefore

$$\mathbb{E}_A(H(Y)) \geq n - \log(\mathbb{E}_A(n_x))$$

But $\mathbb{E}_A(n_x)$ can be computed explicitly as follows:

$$\mathbb{E}_A(n_x) = \mathbb{E}_A\left(\sum_{x' \in \{\pm 1\}^n} 1_{\varphi(x)=\varphi(x')}\right) = \sum_{x' \in \{\pm 1\}^n} \mathbb{P}(\varphi(x) = \varphi(x'))$$

In the last expression $A$ and hence $\varphi$ are random and the probability is computed according to this randomness.

Now note that $\mathbb{P}(\varphi(x) = \varphi(x')) = \mathbb{P}(\varphi(x)_i = \varphi(x')_i \text{ for } i = 1, 2, \ldots, m)$ and since the rows of $A$ are independent, $\varphi(x)_i = \varphi(x')_i$'s are independent events and hence

$$\mathbb{P}(\varphi(x) = \varphi(x')) = \prod_{i=1}^{m} \mathbb{P}(\varphi(x)_i = \varphi(x')_i)$$

But $\mathbb{P}(\varphi(x)_i = \varphi(x')_i) = 0$ if $x$ and $x'$ differ in an odd number of elements and $\mathbb{P}(\varphi(x)_i = \varphi(x')_i) = \frac{\binom{2j}{j}}{2^{2j}}$ if they differ in $2j$ positions. Thus, since there are $\binom{n}{2j}$ different $x'$'s that have $2j$ positions not equal to $x$, we get

$$\sum_{x' \in \{\pm 1\}^n} \mathbb{P}(\varphi(x) = \varphi(x')) = \sum_{j=0}^{\lfloor \frac{n}{2} \rfloor} \binom{n}{2j} \left(\frac{\binom{2j}{j}}{2^{2j}}\right)^m$$

Substituting in the previous inequality, we obtain

$$\mathbb{E}_A(H(Y)) \geq n - \log\left(\sum_{j=0}^{\lfloor \frac{n}{2} \rfloor} \binom{n}{2j} \left(\frac{\binom{2j}{j}}{2^{2j}}\right)^m\right)$$

And since there are always values not less than the expected value, we have

$$C(m,n) = \max_{A} C(A) \geq \max_{\substack{A \\ Y=AX \\ X \text{ uniform}}} H(Y) \geq \mathbb{E}_A\big(H(Y)\big)$$

And thus equation (2) is derived. ∎

### 3.2. Noisy channels

In this section, we will investigate the sum capacity for binary CDMA in the presence of noise. The setting is as before but we assume that there is an additive noise vector $N = [N_1, \ldots, N_m]^T$ such that $N_i$'s are *i.i.d.* random variables with a given pdf $f$. The sum capacity function in this case is defined as

$$C(m,n,f) = \max_{A \in \mathcal{M}_{m \times n}(\pm 1)} C(A,f)$$

where $C(A,f)$ is the sum channel capacity for the CDMA with signature matrix $A$, which is equal to $\max I(X;Y)$ according to Proposition (1), where the maximum is taken over all product distributions on $X$.

**Theorem 2: A general Lower Bound for a Noisy channel:** For any function $q$,

$$C(m,n,f) \geq n - m\mathbb{E}(q(N_1)) - \log\left(\sum_{k=0}^{n} \binom{n}{k}\left(\mathbb{E}\left(2^{-q\left(N_1 - \frac{2S_k}{\sqrt{m}}\right)}\right)\right)^m\right) \quad (3)$$

where $S_k$ is the sum of $k$ independent random variables taking $\pm 1$ with equal probability (also independent of $N_1$).

**Proof:** To prove this theorem, our approach is again to pick a matrix at random and then try to estimate the expected value of the mutual information of the channel corresponding to this matrix. The result is clearly a lower bound since the maximum is always greater than the expected value.

Let us fix a signature matrix $A$ and let $Y = \frac{1}{\sqrt{m}} AX + N$ where $X$ is uniformly distributed in $\{\pm 1\}^n$ (that is $X_i'$s are independent and equal to $+1$ or $-1$ with probability $\frac{1}{2}$).

We start by the following formula for the mutual information:

$$I(X;Y) = -\int f_{X,Y}(x,y) \log\left(\frac{f_X(x) f_Y(y)}{f_{X,Y}(x,y)}\right) dx\, dy = \mathbb{E}_{X,Y}\left(-\log\left(\frac{f_X(X) f_Y(Y)}{f_{X,Y}(X,Y)}\right)\right)$$

But

$$f_X(x) = 2^{-n}, \quad f_Y(y) = 2^{-n} \sum_{u \in \{\pm 1\}^n} f_N\left(y - \frac{1}{\sqrt{m}} Au\right), \quad f_{X,Y}(x,y) = 2^{-n} f_N\left(y - \frac{1}{\sqrt{m}} Ax\right)$$

Since $Y = \frac{1}{\sqrt{m}} AX + N$, we obtain the following relation

$$I(X;Y) = \mathbb{E}_{X,N}\left(-\log\left(\frac{2^{-n} \cdot 2^{-n} \sum_{u \in \{\pm 1\}^n} f_N\left(\frac{1}{\sqrt{m}} A(X-u) + N\right)}{2^{-n} f_N(N)}\right)\right)$$

$$= n - \mathbb{E}_{X,N}\left(\log\left(\frac{\sum_{u \in \{\pm 1\}^n} f_N\left(\frac{1}{\sqrt{m}} A(X-u) + N\right)}{f_N(N)}\right)\right)$$

Now assume that the signature matrix is also chosen at random and take the expectation with respect to this source of randomness:

$$\mathbb{E}_A(I(X;Y)) = n - \mathbb{E}_{X,N,A}\left(\log\left(\frac{\sum_{u \in \{\pm 1\}^n} f_N\left(\frac{1}{\sqrt{m}} A(X-u) + N\right)}{f_N(N)}\right)\right)$$

According to the symmetry of the above expression in vertices of the hypercube $\{\pm 1\}^n$ as the values of $X$, we can remove the expectation with respect to $X$ and set $X = x_0$ where $x_0 \in \{\pm 1\}^n$ is arbitrary:

$$\mathbb{E}_A\big(I(X;Y)\big) = n - \mathbb{E}_{N,A}\left(\log\left(\frac{\sum_{u\in\{\pm 1\}^n} f_N\left(\frac{1}{\sqrt{m}}A(x_0 - u) + N\right)}{f_N(N)}\right)\right)$$

Now let $q$ be an arbitrary function and rewrite the above formula as

$$\mathbb{E}_A\big(I(X;Y)\big) = n - \mathbb{E}_{N,A}\left(\log\left(2^{\sum q(N_i)} \frac{\sum_{u\in\{\pm 1\}^n} f_N\left(\frac{1}{\sqrt{m}}A(x_0 - u) + N\right)}{2^{\sum q(N_i)} f_N(N)}\right)\right)$$

$$= n - \mathbb{E}_N\left(\sum q(N_i)\right) - \mathbb{E}_{N,A}\left(\log\left(\frac{\sum_{u\in\{\pm 1\}^n} f_N\left(\frac{1}{\sqrt{m}}A(x_0 - u) + N\right)}{2^{\sum q(N_i)} f_N(N)}\right)\right)$$

Using the concavity of the logarithm function and applying Jensen's inequality, we have

$$\mathbb{E}_A\big(I(X;Y)\big) \geq n - \mathbb{E}_N\left(\sum q(N_i)\right) - \log\left(\mathbb{E}_{N,A}\left(\frac{\sum_{u\in\{\pm 1\}^n} f_N\left(\frac{1}{\sqrt{m}}A(x_0 - u) + N\right)}{2^{\sum q(N_i)} f_N(N)}\right)\right)$$

Recall that

$$\mathbb{E}_N\left(\sum q(N_i)\right) = m\mathbb{E}(q(N_1))$$

Now since $N_j$'s are i.i.d., we have

$$f_N(x) = \prod_{j=1}^m f_{N_j}(x_j) = \prod_{j=1}^m f(x_j)$$

and hence

$$\frac{f_N\left(\frac{1}{\sqrt{m}}A(\boldsymbol{x}_0-\boldsymbol{u})+N\right)}{2^{\Sigma q(N_i)}f_N(N)} = \prod_{j=1}^{m}\frac{f\left(\frac{1}{\sqrt{m}}A(\boldsymbol{x}_0-\boldsymbol{u})_j+N_j\right)}{2^{q(N_j)}f(N_j)}$$

But $A(\boldsymbol{x}_0-\boldsymbol{u})_j$'s and $N_j$'s are independent for different $j$'s. Thus we have a product of independent terms and consequently

$$\mathbb{E}_{N,A}\left(\frac{f_N\left(\frac{1}{\sqrt{m}}A(\boldsymbol{x}_0-\boldsymbol{u})+N\right)}{2^{\Sigma q(N_i)}f_N(N)}\right) = \prod_{j=1}^{m}\mathbb{E}_{N_j,A}\left(\frac{f\left(\frac{1}{\sqrt{m}}A(\boldsymbol{x}_0-\boldsymbol{u})_j+N_j\right)}{2^{q(N_j)}f(N_j)}\right)$$

$$= \left(\mathbb{E}_{N_1,A}\left(\frac{f\left(\frac{1}{\sqrt{m}}A(\boldsymbol{x}_0-\boldsymbol{u})_1+N_1\right)}{2^{q(N_1)}f(N_1)}\right)\right)^m$$

But

$$\mathbb{E}_{N_1}\left(\frac{f\left(\frac{1}{\sqrt{m}}A(\boldsymbol{x}_0-\boldsymbol{u})_1+N_1\right)}{2^{q(N_1)}f(N_1)}\right) = \int \frac{f\left(\frac{1}{\sqrt{m}}A(\boldsymbol{x}_0-\boldsymbol{u})_1+x\right)}{2^{q(x)}f(x)}f(x)dx$$

$$= \int f\left(\frac{1}{\sqrt{m}}A(\boldsymbol{x}_0-\boldsymbol{u})_1+x\right)2^{-q(x)}dx = \int f(x)2^{-q\left(x-\frac{1}{\sqrt{m}}A(\boldsymbol{x}_0-\boldsymbol{u})_1\right)}dx$$

$$= \mathbb{E}_{N_1}\left(2^{-q\left(N_1-\frac{1}{\sqrt{m}}A(\boldsymbol{x}_0-\boldsymbol{u})_1\right)}\right)$$

Now note that if $\boldsymbol{x}_0$ and $\boldsymbol{u}$ have $k$ different entries, then $\frac{1}{\sqrt{m}}A(\boldsymbol{x}_0-\boldsymbol{u})_1$ has the same distribution as $\frac{2S_k}{\sqrt{m}}$, where $S_k$ is the sum of $k$ independent random variables taking $\pm 1$ with equal probability. Thus the last expression can be written as

$$\left(\mathbb{E}\left(2^{-q\left(N_1-\frac{2S_k}{\sqrt{m}}\right)}\right)\right)^m$$

And hence we obtain

$$\mathbb{E}_A(C(A,f)) \geq \mathbb{E}_A(I(X;Y)) \geq n - m\mathbb{E}(q(N_1)) - \log\left(\sum_{k=0}^{n}\binom{n}{k}\left(\mathbb{E}\left(e^{-q\left(N_1-\frac{2S_k}{\sqrt{m}}\right)}\right)\right)^m\right)$$

Since the uniform distribution is just one choice of product distributions for the input vector, one has

$$C(A,f) \geq I(X;Y)$$

And thus

$$\mathbb{E}_A(C(A,f)) \geq \mathbb{E}_A(I(X;Y))$$

But then

$$C(m,n,f) = \max_A C(A,f) \geq \mathbb{E}_A(C(A,f))$$

This completes the proof. ∎

**Corollary 1:** If we let $q(x) = -\gamma \log f(x)$ where $f$ is the pdf of the additive noise, then from (3) we arrive at the following family of lower bounds

$$C(m,n,f) \geq n - m\gamma(h(f)) - \log\left(\sum_{k=0}^{n}\binom{n}{k}\left(\sum_{j=0}^{k}\frac{\binom{k}{j}}{2^k}g_\gamma\left(\frac{4j-2k}{\sqrt{m}}\right)\right)^m\right) \quad (4)$$

where $h(f) = -\int f(x)\log(f(x))\,dx$ is the differential entropy for the pdf $f$ and the function $g_\gamma$ is defined by $g_\gamma(t) = \int f(t+x)f(x)^\gamma dx$.

**Remark 1:** In both noiseless and noisy cases, our approach is based on a probabilistic argument; we choose a signature matrix at random and then try to estimate the expected value of the sum capacity of the channel corresponding to this random matrix. Consequently, our lower bounds are in fact bounds for the average sum channel capacity of a typical signature matrix. This point combined with the concentration theorems mentioned by Korada and Macris [15]-[16] imply that the sum capacity of a channel with a random signature matrix (for large $n$ and $m$) is greater than the bounds obtained in theorems (1) and (2) with high probability.

**Example 1: Gaussian noise**

An important special case is Additive White Gaussian Noise (AWGN) with variance $\sigma^2$. If we denote the capacity in this case by $C_G(m, n, \sigma^2)$, then we have the following family of lower bounds:

**Proposition 2:** For any positive real number $\gamma$,

$$C_G(m,n,\sigma^2) \geq n - m\gamma \log(\sqrt{e}) - \log\left(\sum_{k=0}^{n} \binom{n}{k} \left(\sum_{j=0}^{k} \frac{\binom{k}{j}}{2^k} \frac{e^{-2\left(\frac{2j-k}{\sigma\sqrt{m}}\right)^2 \left(\frac{\gamma}{1+\gamma}\right)}}{\sqrt{1+\gamma}}\right)^m\right) \quad (5)$$

The proof is rather straightforward from (4). This family of the lower bounds along with its envelope (supremum) is simulated in Figs. 2-4 and will be discussed in Section 5.

**Example 2: Uniform distribution**

Assume that the noise is of the form $N = [N_1, \ldots, N_m]^T$ where $N_i$'s are *i.i.d.* random variables with uniform distribution on $[-a, a]$. Denote the capacity in this case by $C_U(m, n, a)$.

**Proposition 3:**

$$C_U(m,n,a) \geq n - \log\left(\sum_{k=0}^{n} \binom{n}{k} \left(\sum_{j=0}^{k} \frac{\binom{k}{j}}{2^k} \psi\left(\frac{4j-2k}{a\sqrt{m}}\right)\right)^m\right) \quad (6)$$

where the function $\psi$ is defined as

$$\psi(u) = \begin{cases} 1 - |u| & if \ |u| \leq 1 \\ 0 & if \ |u| > 1 \end{cases}$$

The proof is again straightforward from (4). Notice that for the uniformly distributed noise the parameter $\gamma$ disappears due to cancellation.

**Corollary 2:** The noiseless lower bound ((2) from Theorem 1) can be derived from (4) of Corollary 1 when noise goes to zero.

Proof: Let $f$ be an arbitrary pdf and $\varepsilon > 0$. Define $f_\varepsilon(t) = \frac{1}{\varepsilon} f(\frac{t}{\varepsilon})$. It can be easily checked that $h(f_\varepsilon) = h(f) + \log(\varepsilon)$ and $g_\varepsilon^\gamma(t) = \varepsilon^{-\gamma} g_\gamma(\frac{t}{\varepsilon})$ where $g_\varepsilon^\gamma(t) = \int f_\varepsilon(t+x) f_\varepsilon(x)^\gamma dx$ and $g_\gamma(t) = \int f(t+x) f(x)^\gamma dx$ as before. Now, from (4), we know that

$$C(m, n, f_\varepsilon) \geq n - m\gamma(h(f) + \log(\varepsilon)) - \log\left(\sum_{k=0}^{n} \binom{n}{k} \left(\sum_{j=0}^{k} \frac{\binom{k}{j}}{2^k} \varepsilon^{-\gamma} g_\gamma\left(\frac{4j - 2k}{\varepsilon\sqrt{m}}\right)\right)^m\right)$$

$$= n - m\gamma h(f) - \log\left(\sum_{k=0}^{n} \binom{n}{k} \left(\sum_{j=0}^{k} \frac{\binom{k}{j}}{2^k} g_\gamma\left(\frac{4j - 2k}{\varepsilon\sqrt{m}}\right)\right)^m\right).$$

But for fixed values of $j, k$ as $\varepsilon \to 0$, $g_\gamma\left(\frac{4j-2k}{\varepsilon\sqrt{m}}\right) \to 0$ if $2j \neq k$ and $g_\gamma\left(\frac{4j-2k}{\varepsilon\sqrt{m}}\right) \to g_\gamma(0) = \int f(x)^{1+\gamma} dx$ if $2j = k$. Hence, we obtain

$$\lim_{\varepsilon \to 0} C(m, n, f_\varepsilon) \geq n - m\gamma h(f) - \log\left(\sum_{j=0}^{n} \binom{n}{k} \left(\frac{\binom{2j}{j}}{2^{2j}} g_\gamma(0)\right)^m\right).$$

It can be seen that the right-hand side takes its maximum at $\gamma = 0$ since $g_0(0) = \int f(x)\, dx = 1$; thus the noiseless bound shown in (2) is derived. ∎

## 4. A Conjectured Upper Bound for the Sum Channel Capacity

As stated in Proposition 1 in Section 2, the sum capacity of a multiple access channel is the maximum mutual information over all product distributions on the input vector. But the symmetry between users and between $-1$'s and $1$'s in CDMA channels makes it plausible to think that the best probability distribution is the uniform distribution (the symmetric distribution with respect to users) on the input vector. Although this statement seems obvious, it is an open problem (please refer to footnote 6 for more details). In [16], Korada and Macris conjecture that this is the case for the Gaussian white noise. Here, we state the conjecture in an extended form, where the noise can have any general symmetric pdf:

**Conjecture:** Let $A \in \mathcal{M}_{m \times n}(\pm 1)$ and $Y = \frac{1}{\sqrt{m}} AX + N$, where $N = [N_1, \dots, N_m]^T$ constitutes *i.i.d.* random variables with a given symmetric pdf $f$ (i.e., $f(x) = f(-x)$). Then $\max_{\prod p_i(x_i)} I(X;Y)$ is attained for the uniform distribution when $X_i$'s are independent and equal to $\pm 1$ with probability $\frac{1}{2}$.

It is not difficult, as shown by Korada and Macris, to see that if we consider the average mutual information over all signature matrices, then the maximum is attained for the uniform distribution [16], but the problem for a specific channel comes to be much more difficult.

In this section, assuming this conjecture is true, we will derive a conjectured upper bound for the sum capacity, which is very close to the lower bound obtained in the previous section. This implies that the bounds are relatively tight.

**Proposition 4 Conjectured Upper Bound:** Let $f$ be a symmetric probability distribution function, that is $f(x) = f(-x)$. Defining the function $\tilde{f}$ by $\tilde{f}(x) = \sum_{j=0}^{n} \frac{\binom{n}{j}}{2^n} f\left(x - \frac{2j-n}{\sqrt{m}}\right)$, we have

$$C(m, n, f) \leq \min\left(n, m\left(h(\tilde{f}) - h(f)\right)\right) \qquad (7)$$

where $h(f)$ is the differential entropy as defined in (4). For the noiseless case, we should use the usual entropy instead of the differential entropy as will be shown in Example 5.

**Proof:** It is clear that for binary inputs $C(m, n, f) \leq n$. Assume that $\boldsymbol{A}$ is the signature matrix that results in the maximum mutual information between $X$ and $Y = \frac{1}{\sqrt{m}} \boldsymbol{A} X + N$. Based on our conjecture, for any signature matrix including $\boldsymbol{A}$, the mutual information is maximized at the uniform distribution for the input vector. Thus by assuming that $X_i$'s are independent and uniform, we have

$$C(m, n, f) = C(\boldsymbol{A}, f) = I(X; Y) = h(Y) - h(Y|X)$$

But

$$h(Y|X) = h(N) = m\, h(f)$$

and

$$h(Y) = h(Y_1, \ldots, Y_m) \leq \sum_{j=1}^{m} h(Y_j)$$

Now note that $Y_j = \frac{1}{\sqrt{m}} \sum_i a_{ji} X_i + N_j$, where $\sum_i a_{ji} X_i$ is the sum of independent symmetric Bernoulli random variables taking $\pm 1$. Hence $Y_j$ has the distribution of a convolved binomial random variable and a random variable with pdf $f$ which is exactly $\tilde{f}$. By Substituting in the above relations, the proof is complete. ∎

**Example 3 (The Gaussian noise):**

For a Gaussian distribution, the function $\tilde{f}$ becomes

$$\tilde{f}(x) = \frac{1}{\sigma\sqrt{2\pi}} \sum_{j=0}^{n} \frac{\binom{n}{j}}{2^n} e^{-\frac{\left(x - \frac{2j-n}{\sqrt{m}}\right)^2}{2\sigma^2}}$$

Then, we have

$$C_G(m, n, \sigma^2) \leq \min\left(n, m\left(h(\tilde{f}) - \log(\sqrt{2\pi e}\sigma)\right)\right)$$

**Example 4 (The noise with uniform distribution):**

For a uniform distribution, the function $\tilde{f}$ becomes

$$\tilde{f}(x) = \frac{1}{2a} \sum_{j=0}^{n} \frac{\binom{n}{j}}{2^n} 1_{[-a,a]}\left(x - \frac{2j-n}{\sqrt{m}}\right)$$

Then, we have

$$C_U(m, n, a) \leq \min\left(n, m(h(\tilde{f}) - \log(2a))\right)$$

Reference [29] also discusses the maximum entropy of the sum of independent random variables.

**Example 5 (The noiseless channel):**

For the noiseless case, we can assume that the noise pdf is an impulse. Consequently, $\tilde{f}$ becomes

$$\tilde{f}(x) = \sum_{j=0}^{n} \frac{\binom{n}{j}}{2^n} \delta\left(x - \frac{2j-n}{\sqrt{m}}\right)$$

This is a discrete probability distribution. Hence we should use the usual entropy instead of the differential entropy and a true upper bound as opposed to the conjectured[6] one shown in (7) becomes

---

[6] Based on the book by Marshall and Olkin [27], for binary inputs with equal probability, the upper bound conjecture is a true bound for the noiseless case. Apparently, Chang & Weldon [24] has assumed this without a proper reference in their derivation. Tanaka, Korada & Macris ([15-16]) have conjectured that mutual information is maximized for symmetric binary inputs with additive Gaussian noise. Our conjecture for the noisy case with symmetric pdfs is still an open problem. Thus, we have referred everywhere to our upper bound as "conjectured upper bound" for the symmetric noisy cases; in the noisy cases such as Examples 3 and 4, an obvious true upper bound is the noiseless case in Example 5.

$$C(m,n) \leq \min\left(n, \left(m\, H(\tilde{f})\right)\right)$$

This bound is identical to that of Chang and Weldon for T-user MAC's [24].

## 5. Asymptotic Analysis

In order to compare our results to Tanaka's sum capacity bound as shown below [12]-[16], and [26], we first try to find the normalized lower bound for the sum capacity in the limit when $n$ and $m$ go to infinity while keeping the ratio $\beta = n/m$ constant. Subsequently, the same limit is found for the conjectured upper bound. Simulation results in Section 6 show a comparison of our bounds in the limit with that of Tanaka's bound.

### 5.1. Lower Bounds for the Sum Channel Capacity in the Limit

It can be seen from [17] and [30] that if $\sigma = 0$, for any $\beta$, $\lim_{\substack{n/m=\beta \\ n,m\to\infty}} c(m,n,f) = 1$. This is because $n_{th}$ (the maximum number of users without any interference defined in [30]) grows faster than linear with respect to $m$. In the following theorem, we represent a limiting procedure which we claim is the appropriate regime for the noiseless case and derive the exact capacity in the limit in Theorems 3 and 5.

**Theorem 3 Noiseless case**: If $c(m,n)$ denotes the maximum capacity per user, i.e., $c(m,n) = \frac{1}{n}C(m,n)$, then for any $\zeta$

$$\lim_{\substack{n/(m\log n)\to\zeta \\ n,m\to\infty}} c(m,n) \geq \min\left\{1, \frac{1}{2\zeta}\right\} \quad (8)$$

The proof is given in Appendix A.

**Therem 4**: (**Noisy channel**): For any arbitrary function $q(x)$,

$$\lim_{\substack{n/m=\beta \\ n,m\to\infty}} c(m,n,f) \geq 1 - \sup_{t\in[0,1]} \left[H(t) + \frac{1}{\beta}\left(\mathbb{E}(q(N_1)) + \log \mathbb{E}\left(2^{-q(N_1 - 2\sqrt{t\beta}Z)}\right)\right)\right]$$

Considering the family $q(x) = -\gamma \log f(x)$, we obtain the following bound for the Gaussian case:

**Special Case (Gaussian Noise):**

$$\lim_{\substack{n/m=\beta \\ n,m\to\infty}} c(m,n,\sigma^2) \geq 1 - \inf_{\gamma} \sup_{t\in[0,1]} \left[H(t) + \frac{1}{2\beta}\left(\gamma\log e - \log\left(1 + \frac{\gamma}{\sigma^2}(\sigma^2 + 4t\beta)\right)\right)\right] \quad (9)$$

The proof is given in Appendix B.

Although (9) has been derived by just considering a family of special functions for $q(x)$, we will prove in Appendix D that the best bound obtainable by any $q(x)$ cannot be much better than (9). Moreover, our simulation results show that the formula (D1) in Appendix D can be used as a good approximation for (9), which is computationally much simpler.

### 5.2. Conjectured Upper Bounds for the Sum Channel Capacity in the Limit

**Theorem 5: Noiseless Channel:** In the limit, we have

$$\lim_{\substack{n/(m\log n)=\zeta \\ n,m\to\infty}} c(m,n) \leq \min\left\{1, \frac{1}{2\zeta}\right\} \quad (10)$$

**Proof:** From Example 5 and [24], we have

$$\lim_{\substack{n/(m\log n)=\zeta \\ n,m\to\infty}} c(m,n) = \min(1, \frac{m\, H(\tilde{f})}{n = \zeta m \log n}) \leq \min(1, \frac{\log(2\pi en)}{2\zeta \log n}) \to \min\left\{1, \frac{1}{2\zeta}\right\} \quad \blacksquare$$

**Theorem 6**: **Noisy Channel**: In the limit, we have

$$\lim_{\substack{n/m=\beta \\ n,m\to\infty}} c(m,n,f) \leq \min\left\{1, \frac{1}{\beta}\left(h(N_1 + \sqrt{\beta}Z) - h(N_1)\right)\right\} \quad (11)$$

where $Z$ is a standard Gaussian random variable independent of $N_1$.

**Proof:** The $f$ obtained from (7) of Proposition 4 is equal to the pdf of the r.v.

$$N_1 + \frac{S_n}{\sqrt{m}} = N_1 + \sqrt{\beta}\frac{S_n}{\sqrt{n}}.$$

Due to the central limit theorem, the term on the right hand side approaches a Gaussian r.v. in the limit and thus we have (11). ∎

**Example**: for the Gaussian noise, when $N_1$ is a Gaussian random variable of variance $\sigma^2$, one has $h(N_1) = \frac{1}{2}\log(2\pi e\sigma^2)$ and $h(N_1 + \sqrt{\beta}Z) = \frac{1}{2}\log(2\pi e(\sigma^2 + \beta))$. Hence

$$\lim_{\substack{n/m=\beta \\ n,m\to\infty}} c(m,n,f) \leq \min\left\{1, \frac{1}{2\beta}\log\left(1 + \frac{\beta}{\sigma^2}\right)\right\} \quad (12)$$

The above upper bound is reminiscent of the Shannon capacity for an AWGN channel ($1/\beta = m/n$ is the normalized bandwidth and SNR $= \frac{\beta}{\sigma^2}$). Theorems 3 and 5 for the noiseless cases show that the ratio of the upper and lower bounds approach 1. However, there is always a gap between the bounds for finite $m$ and $n$. When $\beta$ approaches 0, the above bound approaches $\frac{\log(e)}{2\sigma^2}$ which is not a good bound for low SNR values since only for $E_b/N_0 = \frac{1}{2\sigma^2} \leq -1.593 dB$, the above bound is less than 1 bit/user (see Fig. 8). However, for $\beta \leq 1$, we have the actual channel capacity, which is equal to the single user capacity.

## 6. Simulation Results

The lower and conjectured upper bounds have been simulated with respect to $n$, $m$ and $E_b/N_0$. Also, the asymptotic bounds are simulated and compared to the results of Tanaka. For the noiseless case, simulations of the normalized lower and upper bounds derived from Theorem 1 and Example 5 of Proposition 4 have been extensively studied in [30] and are given in Fig. 1 as a reference. This figure shows that there exist CDMA matrices $A$ such that the number of users, $n$, can be 3-4 times the spreading gain, $m$. For example, for $m = 64$, the maximum number of users with almost no interference is close to $n = 239$. For $n \leq 64$, orthogonal Walsh matrices can be used, and for $64 < n < 239$, there are matrices [30] that are interference free.

For the AWGN case, according to (5) from Proposition (2), a family of lower bounds for the normalized sum channel capacity for any number $\gamma$ can be obrtained. Fig. 2 shows some of the bounds of this family for various values of $\gamma$ with respect to $n$. Also the envelope (supremum) of the whole family is plotted in the same figure when $E_b/N_0$ is fixed[7] and is equal to 16 dB. This envelope behaves similarly to the noiseless case of Fig.1, and as expected, the sum capacity is reduced. Fig. 3 shows the normalized lower and conjectured upper bounds for two values of $E_b/N_0$. This figure shows that the noisier the channel, the tighter is our lower and conjectured upper bounds.

The variation of the normalized sum capacity bounds w.r.t. $E_b/N_0$ for various number of users $n$ is shown in Fig. 4. This figure shows the logarithmic relationship between the capacity and the signal-to-noise-ratio. The reason of the leveling off of the curves is due to the fact that the normalized sum capcity cannot be greater than one. This figure shows that for larger $n$ given a fixed $m$ (i.e., larger $\beta$) the bound is tighter when $E_b/N_0$ is low. However, for higher $E_b/N_0$ and the smaller $n$ given a fixed $m$, the upper and lower bounds will approach 1 bit/user faster and hence are tighter in that region.

In summary, the sum capacity bounds behave differently for different $E_b/N_0$ values and spreading gains. In general, for a given $E_b/N_0$ and $m$, the sum capacity bounds are very tight and increase almost linearly with $n$ (1 bit/user) up to a point (this point increases with increasing $m$ and $E_b/N_0$) and then, suddenly, the interfereneces of users dominate as $n$ is increased. Beyond this point, the conjectured upper and lower bounds are no longer as tight. However, the higher $\beta$ and $E_b/N_0$, the tighter are the bounds in this region.

### 6.1. Comparison with Tanaka's bound in the limiting case

---

[7] $E_b/N_0$ for CDMA is defined as $1/2\sigma^2$ assuming that each user is normalized to $\pm 1$. The signature matrix is assumed to be normalized by $1/\sqrt{m}$. The chip period is also normalized to one unit of time.

Although Tanaka's definition of the sum capacity is slightly different from ours[8], we would like to compare our bounds, derived from the limiting case in Section 5, with that of Tanaka's. The reader should bear in mind that our results (even for the asymptotic cases) are valid for any additive noise with a general pdf, while Tanaka's asymptotic results are only valid for additive Gaussian noise. Also, Tanaka's results may not be valid for finite dimensional systems, in general.

We have also managed to simulate the complicated formulae of Tanaka[9] from a combination of references [12]-[16] for $\beta = 0.5, 1, 2, 4,$ and $8$. For $\beta \leq 1$, fortunately, we do have the exact capacity since we can use Walsh signature matrices, and due to its orthogonality, its performance is equivalent to binary PSK. The actual capacity is $1 - H(p)$, where $p$ is the probability of error and is related to $E_b/N_0$ through the normal distribution. Figs. 5 and 6 show a comparison of the actual normalized sum capacity with Tanaka's bound for $\beta = 0.5$ and $1$, respectively. Clearly, Tanaka's bound is an upper bound. In these two figures, we have also included our lower and upper bounds. Tanaka's bound becomes tighter when $\beta$ increases from 0.5 to 1; our upper bound also becomes tighter with increasing $\beta$ but is not as good as Tanaka's bound for $E_b/N_0$ values less than 8 dB. As $\beta$ becomes greater than 1, our bounds become tighter, but we have no way to evaluate Tanaka's bound since the actual sum capacity is not known. However, as shown in Fig. 7, Tanaka's bound approaches our upper bound with increasing $\beta$; we can show this analytically from Appendix C.

An important observation from Figs. 5-7 is that that for any $\beta$, the normalized sum capacity can approach 1 bit/user for large values of SNR but this is not true for finite dimensions (due to overloaded interference), and therefore cannot be extrapolated from infinite dimension to finite dimensions.

---

[8] Actually, the definition of sum capacity by Tanaka and references [15-16] is not exactly the same as ours. They maximize, over the input probabilities, the average mutual information w.r.t. the random signature matrices. We define the capacity as the maximum over all matrices and input probabilities. The averaging over all matrices is a lower bound for our definition of the sum capacity.

[9] For the Matlab and Mathcad codes of Tanaka's formulae as well as ours, please check http://acri.sharif.edu/en/g7/kotob/default.asp?page=1

An interesting point about our asymptotic (large scale systems) bounds is that they are very good approximations for finite dimensional CDMA systems. Fig. 8 shows a comparison between lower and conjectured upper bounds for finite dimensional and the asymptotic cases when $\beta = 2$. For $m \geq 4$, the asymptotic upper bound and the finite dimensional one coinside. Also, for $m \geq 32$, there is very little difference between the asymptotic and the finite lower bounds; our computer simulations show that this conclusion is valid for all values of $\beta$ provided that $m$ is adjusted according to the vlaues of $\beta$. Since simulations of the bounds in the limit are much simpler, we can use the lower and upper bounds derived in Section 5- (9)[10] and (12)- instead of the bounds derived from the combinatorics in (5) and (7).

Fig. 9 shows another interesting comparison between the actual bounds developed from combinatorics and the asymptotic bounds extrapolated to finite scales ($m = 8, 64$ and $E_b/N_0 = 16$ dB). For $m = 64$, the two conjectured upper bounds coincide while Tanaka's extrapolated bound is slightly lower than our conjectured upper bound. The exrapolated lower bound is higher than the actual bound derived from combinatorics. Since $\beta = n/64$, as $\beta$ increases, the gap between our upper bound and Tanaka's bound diminishes. For $m = 8$, Fig. 9 shows that our asympotic bounds as well as Tanaka's bound are above our combinatorics upper bound for large number of users ($n \approx 30, i.e., \beta \approx 4$) and therefore not accurate.

This senario changes with higher noise variances. For a high noisy environment and with the previous parameters, Fig. 10 shows that the extrapolated and actual bounds coincide; Tanaka's bound, as usual, is in between closer to the conjectured upper bound. The reason that the bounds coincide can be explained from the bahvior of (5) and (9) for the lower bounds and Example 3 and (12) for the upper bounds when the noise variance goes to infinity. We can easily show that the ratio of (5) and (9) goes to 1 in the limit, and for the conjectured upper bounds, we can show that for large variances, the pdf $\tilde{f}$ in Example 3 becomes Gaussian like with a variance of $\beta + \sigma^2$ (convolution of binomial and Gaussian pdf's). Thus, the upper bound in Example 3 degnerates into (12) and we can show that the ratio goes to 1 in the limit.

---

[10] Eq. (D1) in Appendix D is even a simpler approximation of (5) and (9).

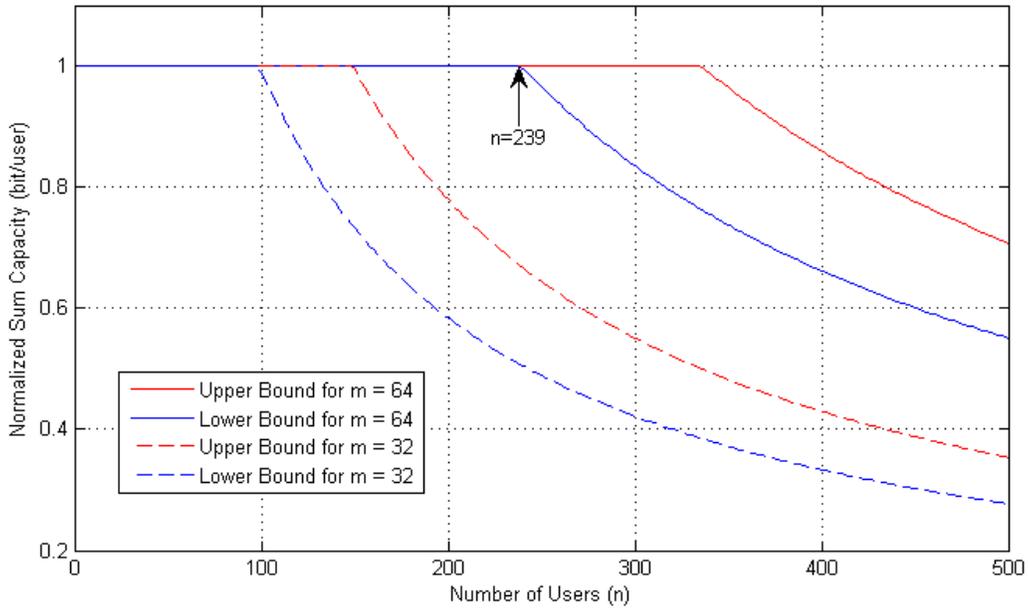

Fig. 1. The lower and upper bounds for the normalized channel capacity vs. the number of users $n$ for different spreading gains $m$ in a noiseless system according to Theorem 1.

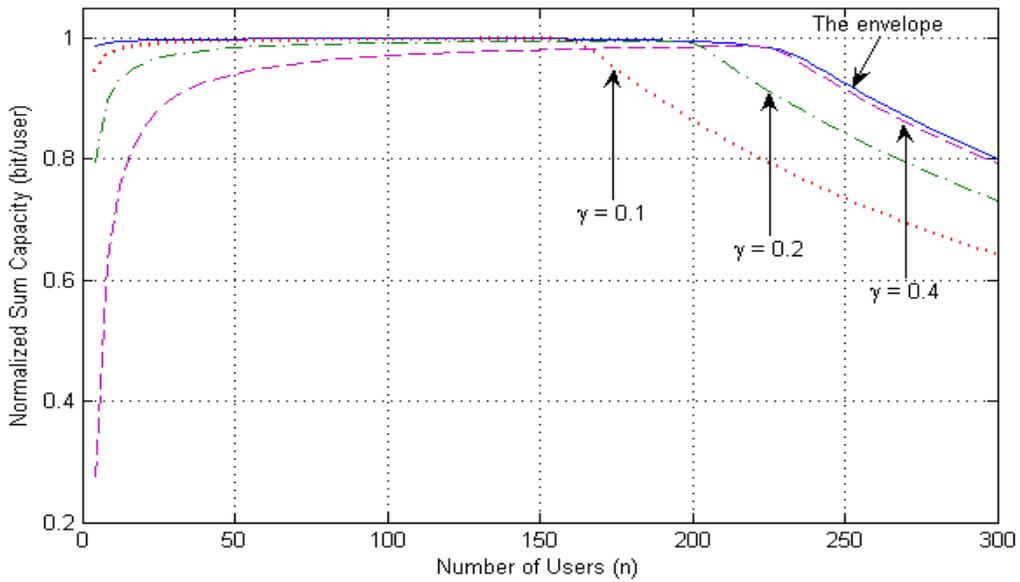

Fig. 2. A family of lower bounds for the sum capacity of an AWGN channel using different $\gamma$'s and their envelope vs. number of users $n$ when $E_b/N_0 = 8$ dB and the spreading gain is $m = 64$.

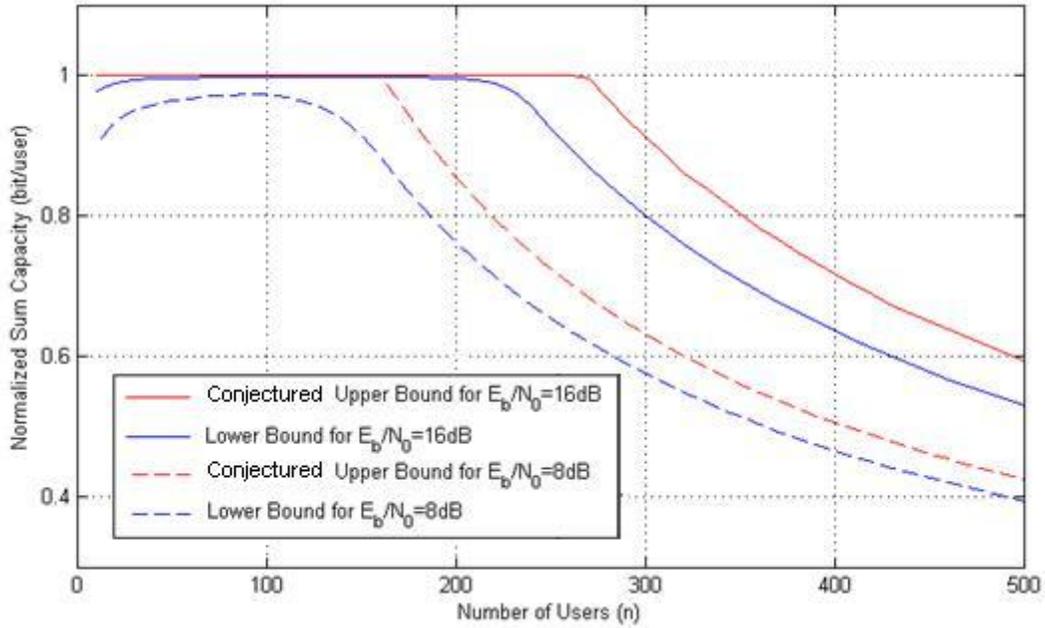

Fig. 3. The normalized sum capacity lower (envelopes) and conjectured upper bounds for AWGN channels vs. $n$ for different $E_b/N_0$ values when $m = 64$.

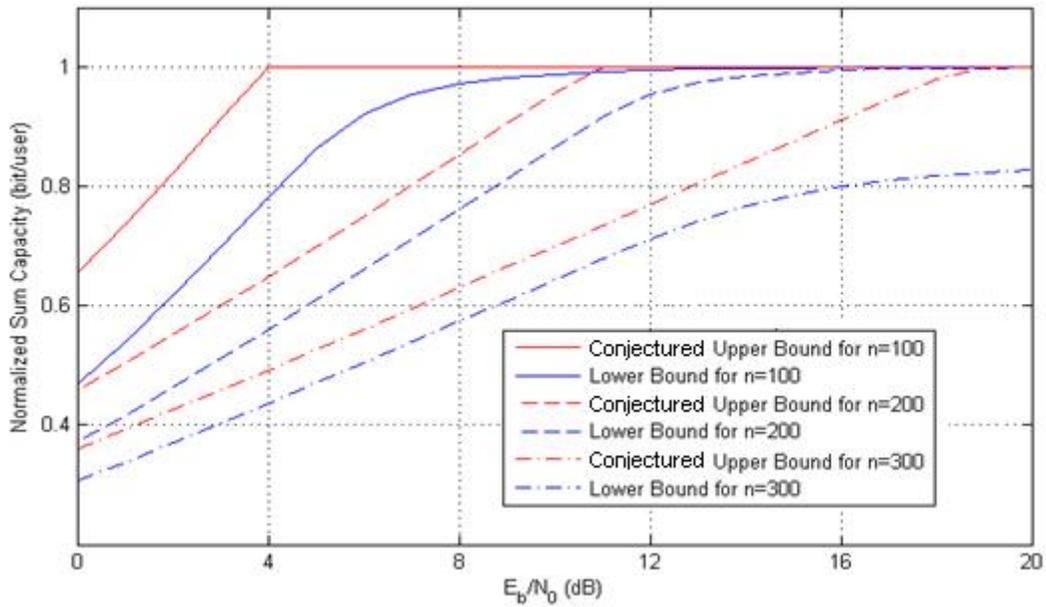

Fig. 4. The normalized sum capacity lower (envelopes) and conjectured upper bounds vs. $E_b/N_0$ for different values of $n$ with $m = 64$.

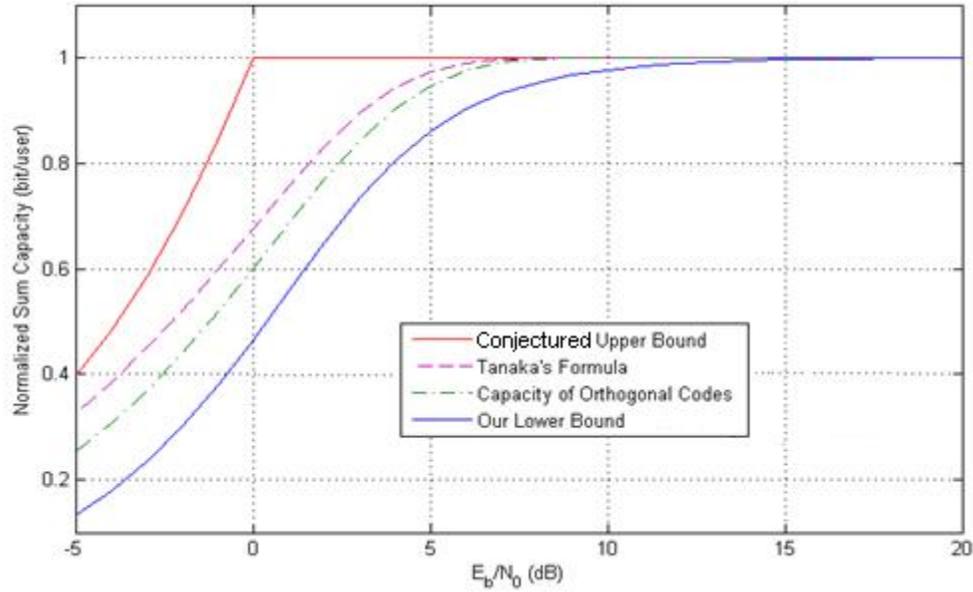

Fig. 5. The normalized sum capacity bounds vs. $E_b/N_0$ in the limit when $n$ and $m$ go to infinity for $\beta = 0.5$ ($m = n$). In this case the orthogonal Walsh signature matrix is equivalent to the single user PSK. Tanaka's result is an upper bound to the actual capacity in this case.

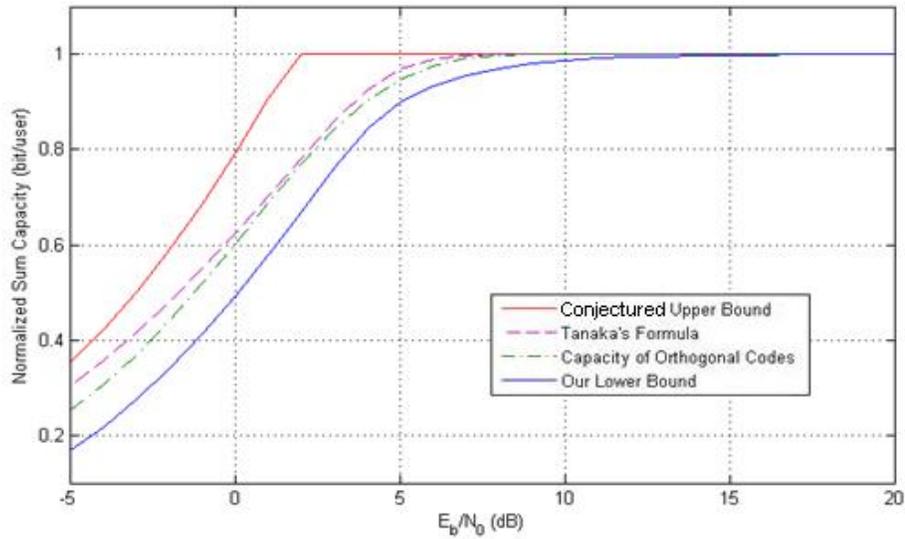

Fig. 6. The normalized sum capacity bounds vs. $E_b/N_0$ in the limit when $n$ and $m$ go to infinity for $\beta = 1$ ($m = n$). In this case the orthogonal Walsh signature matrix is equivalent to the single user PSK. Tanaka's result is a very tight upper bound to the actual capacity in this case.

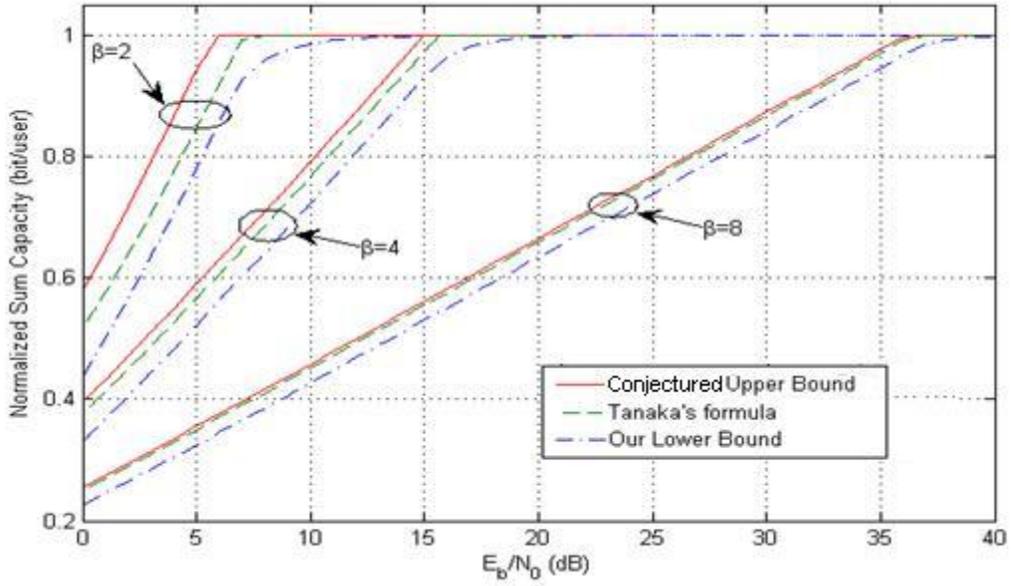

Fig. 7. The normalized sum capacity bounds vs. $E_b/N_0$ in the limit when $n$ and $m$ go to infinity for $\beta = 2, 4$ and $8$. Depending on the values of $\beta$, Tanaka's bound is somewhere between our bounds but closer to our conjectured upper bound as $\beta$ increases. As $\beta$ increases, our lower and conjectured upper bounds and Tanaka's bound become very tight.

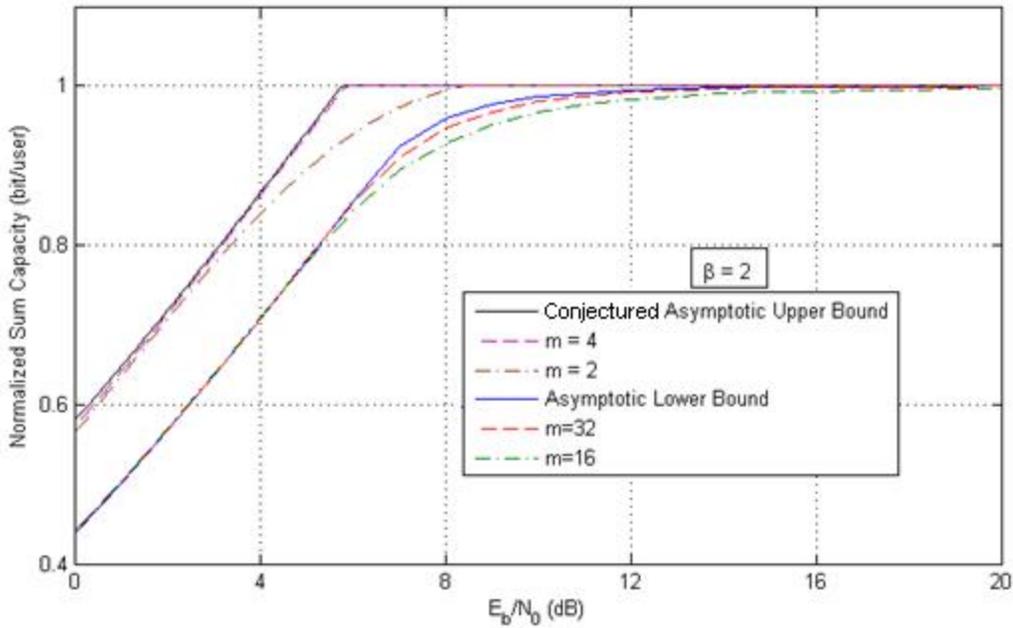

Fig. 8. The normalized sum capacity vs. $E_b/N_0$. As $n$ inclreases with $\beta = 2$, the sum capacity converges to the theoretical limit. For $m \geq 32$, there is very little difference between the liming case and the simulations of finite CDMA.

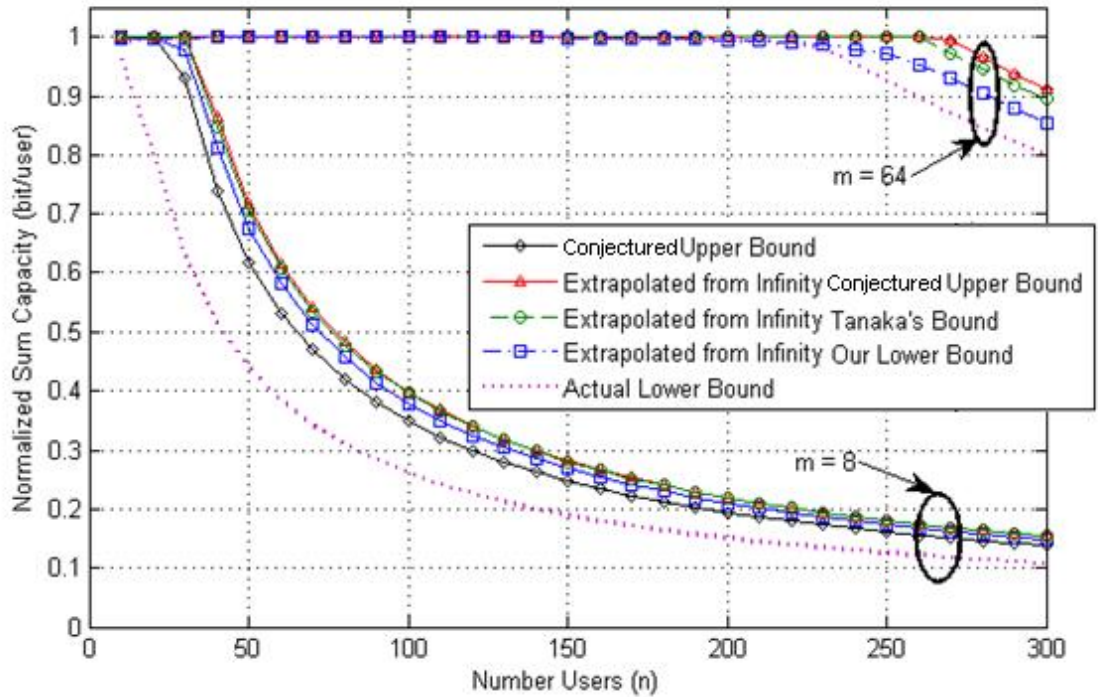

Fig. 9. Comparison of our asymptotic and Tanaka's capacity bounds vs. $n$ extrapolated to finite size scale, $m = 8,\ 64$ and $E_b/N_0 = 16$ dB.

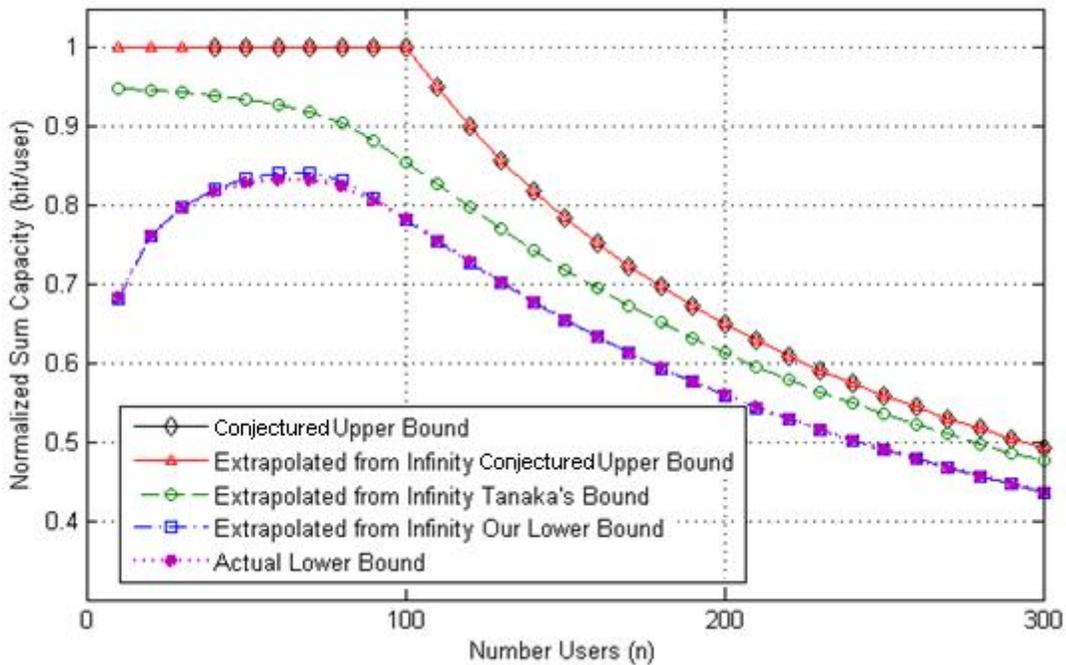

Fig. 10. Comparison of our asymptotic and Tanaka's capacity bounds vs. $n$ extrapolated to finite size scale for a high noisy environment, $m = 64$ and $E_b/N_0 = 4$ dB.

## 7. Conclusion

In this paper, for binary input and binary CDMA signatures, we have derived a family of lower bounds for additive *i.i.d.* noise of any distribution with respect to the number of users, $n$, the spreading gain, $m$, and the pdf of the noise. The envelope of this family gives a lower bound. Special case of AWGN has also been derived and simulated. The bound and simulations show that when the noise variance goes to zero, the noisy and the noiseless lower bounds become identical. We have also derived a conjectured upper bound for a general *i.i.d.* symmetric noise distribution; special cases of Gaussian and uniform distribution are also obtained.

Simulations also show that for a given noise level ($E_b/N_0$) and spreading gain, $m$, the number of users, $n$, can be greater than $m$ with almost no interference, i.e., overloaded CDMA is possible and thus we can improve the throughput of a CDMA system by designing structured CDMA codes. In general, for a given $E_b/N_0$ and $m$, from Figs. 1, 3, and 4, the sum capacity bounds are very tight and increase almost linearly with $n$ ($\approx$ 1 bit/user) up to a point (this point increases with increasing $m$ and $E_b/N_0$) and then, abruptly, the interferenece of users dominates as $n$ is increased. Beyond this point the upper and lower bounds are no longer as tight. However, the higher $\beta$ and $E_b/N_0$, the tighter are the bounds in this region.

We have also simulated Tanaka's bound and showed that it is a tight upper bound in the limiting case for $\beta \leq 1$. As $\beta$ increases, we have shown analytically as well as by simulations that this bound approaches our conjectured upper bound for a wide range of $E_b/N_0$.

For future work, we intend to generalize our bounds to nonbinary user signals and nonbinary CDMA signature matrices. The proof of the conjecture is also another interesting problem to solve. Also, we would like to extend our results to the asynchronous case with near/far effects and imperfect channel estimations and training. Extension to fading channels is another issue to be investigated. The case when additive noise is not *i.i.d.*, is another interesting topic to be investigated.

**Appendix A**

**Proof of Theorem 3 (Section 5.1):** One can easily check that $\frac{\binom{2j}{j}}{2^{2j}} \sim \frac{1}{\sqrt{\pi j}}$. Hence from (2), we have

$$\sum_{j=0}^{\lfloor \frac{n}{2} \rfloor} \binom{n}{2j} \left(\frac{\binom{2j}{j}}{2^{2j}}\right)^m \sim \int_1^{\frac{n}{2}} \binom{n}{2j} \left(\frac{1}{\sqrt{\pi j}}\right)^m dj$$

By using a change of variable $t = \frac{2j}{n}$, the last integral becomes

$$\int_{\frac{2}{n}}^{1} \binom{n}{tn} \left(\frac{\sqrt{2}}{\sqrt{\pi t n}}\right)^m \frac{n}{2} dt$$

Stirling's estimation, $n! \sim \sqrt{2\pi n} \left(\frac{n}{e}\right)^n$, implies that

$$\binom{n}{tn} \sim 2^{n(-t\log(t)-(1-t)\log(1-t))} = 2^{nH(t)}$$

And so we obtain

$$\lim_{\substack{n/(m\log n)\to \zeta \\ n,m\to\infty}} c(m,n) \geq 1 - \lim_{\substack{n/(m\log n)=\zeta \\ n,m\to\infty}} \frac{1}{n}\log\left(\int_{\frac{2}{n}}^{1} \frac{n}{2}\binom{n}{tn}\left(\frac{\sqrt{2}}{\sqrt{\pi tn}}\right)^m dt\right)$$

$$= 1 - \lim_{\substack{n/(m\log n)\to \zeta \\ n,m\to\infty}} \frac{1}{n}\log\left(\int_{\frac{2}{n}}^{1} 2^{nH(t)} \frac{n}{2}\left(\frac{\sqrt{2}}{\sqrt{\pi tn}}\right)^m dt\right)$$

$$= 1 - \lim_{n\to\infty} \frac{1}{n}\log\left(\int_0^1 2^{nh(t)} P_n(dt) dt\right)$$

where $\{P_n(dt)\}$ is a sequence of finite measures on [0,1] given by $\frac{n}{2}\left(\frac{\sqrt{2}}{\sqrt{\pi tn}}\right)^m dt = \frac{n}{2}\left(\frac{\sqrt{2}}{\sqrt{\pi tn}}\right)^{\frac{n}{\zeta \log n}} dt$ for $t \geq \frac{2}{n}$ and 0 for $0 \leq t < \frac{2}{n}$.

But according to Varadhan's lemma [25],

$$\lim_{n\to\infty} \frac{1}{n}\log\left(\int_0^1 2^{nH(t)} P_n(dt)dt\right) = \max_{t\in[0,1]}[H(t) - I(t)] \quad (A1)$$

where $I$ is the rate function which is defined as the unique lower semi-continuous function with the following properties:

(i) $\limsup_{n\to\infty} \frac{1}{n}\log P_n(O) \leq -\inf_{x\in O} I(x)$ for any open $O$.

(ii) $\liminf_{n\to\infty} \frac{1}{n}\log P_n(C) \geq -\inf_{x\in C} I(x)$ for any closed $C$.

Now let $C = [a, b]$. We will distinguish two cases, $a > 0$ and $a = 0$. In the first case, we have

$$\liminf_{n\to\infty} \frac{1}{n} \log P_n(C) = \liminf_{n\to\infty} \frac{1}{n} \log \int_a^b \frac{n}{2}\left(\frac{\sqrt{2}}{\sqrt{\pi t n}}\right)^{\frac{n}{\zeta \log n}} dt$$

$$= \liminf_{n\to\infty} \frac{1}{n}\left(-\frac{1}{2}\frac{n}{\zeta \log n\pi}\log n + \log n - \frac{1}{2} + \log \int_a^b \left(\frac{1}{\sqrt{t}}\right)^{\frac{n}{\zeta \log n}} dt\right)$$

$$= -\frac{1}{2\zeta} + \liminf_{n\to\infty} \frac{1}{n}\left(\log \frac{1}{\frac{-n}{2\zeta \log n}+1} t^{\frac{-n}{2\zeta \log n}+1}\Big|_a^b\right)$$

$$= -\frac{1}{2\zeta} + \liminf_{n\to\infty} \frac{1}{n}(-) + \liminf_{n\to\infty} \frac{1}{n}\left(\log a^{\frac{-n}{2\zeta \log n}+1}\right) = -\frac{1}{2\zeta}$$

If $a = 0$, then

$$\liminf_{n\to\infty} \frac{1}{n} \log P_n(C) = \liminf_{n\to\infty} \frac{1}{n} \log \int_{\frac{1}{n}}^b \left(\frac{1}{\sqrt{tn}}\right)^{\frac{n}{\zeta \log n}} dt$$

$$= \liminf_{n\to\infty} \frac{1}{n}\left(-\frac{1}{2}\frac{n}{\zeta \log n}\log n + \log \int_{\frac{1}{n}}^b \left(\frac{1}{\sqrt{t}}\right)^{\frac{n}{\zeta \log n}} dt\right)$$

$$= -\frac{1}{2\zeta} + \liminf_{n\to\infty} \frac{1}{n}\left(\log \frac{1}{\frac{-n}{2\zeta \log n}+1} t^{\frac{-n}{2\zeta \log n}+1}\Big|_{\frac{1}{n}}^b\right)$$

$$= -\frac{1}{2\zeta} + \liminf_{n\to\infty} \frac{1}{n}\left(\log \left(\frac{1}{n}\right)^{\frac{-n}{2\zeta \log n}+1}\right) = -\frac{1}{2\zeta} + \frac{1}{2\zeta} = 0$$

Thus we should have $I(0) = 0$ and $I(x) = \frac{1}{2\zeta}$ for $x > 0$. Hence from (A1), we have

$$\lim_{n\to\infty} \frac{1}{n} \log \left(\int_0^1 2^{nh(t)} P_n(dt) dt\right) = \max\left\{0, 1 - \frac{1}{2\zeta}\right\} \blacksquare$$

**Appendix B**

**Proof of Theorem 4: Noisy Lower-bound in the limit:** According to (3),

$$c(m, n, f) = \frac{1}{n} C(m, n, f) \geq 1 - \frac{m}{n} \mathbb{E}(q(N_1)) - \frac{1}{n} \log \left(\sum_{k=0}^n \binom{n}{k}\left(\mathbb{E}\left(2^{-q\left(N_1 - \frac{2S_k}{\sqrt{m}}\right)}\right)\right)^m\right)$$

$$\lim_{\substack{n/m=\beta \\ n,m\to\infty}} c(m,n,f) \geq 1 - \frac{1}{\beta}\mathbb{E}(q(N_1)) - \frac{1}{n}\log\left(\sum_{k=0}^{n}\binom{n}{k}\left(\mathbb{E}\left(2^{-q\left(N_1-\frac{2S_k}{\sqrt{m}}\right)}\right)\right)^m\right)$$

$$\sum_{k=0}^{n}\binom{n}{k}\left(\mathbb{E}\left(2^{-q\left(N_1-\left(\frac{2S_k}{\sqrt{m}}\right)\right)}\right)\right)^m \sim \int_0^n \binom{n}{k}\left(\mathbb{E}\left(2^{-q\left(N_1-\frac{2S_k}{\sqrt{m}}\right)}\right)\right)^m dk$$

$$= \int_0^1 \binom{n}{tn}\left(\mathbb{E}\left(2^{-q\left(N_1-\frac{2S_{tn}}{\sqrt{m}}\right)}\right)\right)^m n\,dt$$

Again we have

$$\binom{n}{tn} \sim 2^{nH(t)}$$

And the central limit theorem implies that

$$\mathbb{E}\left(2^{-q\left(N_1-\frac{2S_{tn}}{\sqrt{m}}\right)}\right) = \mathbb{E}\left(2^{-q\left(N_1-2\sqrt{t\beta}\left(\frac{S_{tn}}{\sqrt{tn}}\right)\right)}\right) \to \mathbb{E}\left(2^{-q(N_1-2\sqrt{t\beta}Z)}\right)$$

where $Z \sim N(0,1)$ is a standard Gaussian random variable, independent of $N_1$.

$$\lim_{\substack{n/m=\beta \\ n,m\to\infty}} c(m,n,f) \geq 1 - \frac{1}{\beta}\mathbb{E}(q(N_1)) - \frac{1}{n}\log\left(\int_0^1 \binom{n}{tn}\left(\mathbb{E}\left(2^{-q\left(N_1-\frac{2S_{tn}}{\sqrt{m}}\right)}\right)\right)^m dt\right)$$

$$= 1 - \frac{1}{\beta}\mathbb{E}(q(N_1)) - \frac{1}{n}\log\left(\int_0^1 2^{nH(t)}\left(\mathbb{E}\left(2^{-q(N_1-2\sqrt{t\beta}Z)}\right)\right)^m dt\right)$$

$$= 1 - \frac{1}{\beta}\mathbb{E}(q(N_1)) - \frac{1}{n}\log\left(\int_0^1 2^{n\left[H(t)+\frac{1}{\beta}\log\mathbb{E}\left(2^{-q(N_1-2\sqrt{t\beta}Z)}\right)\right]} dt\right)$$

$$\to 1 - \frac{1}{\beta}\mathbb{E}(q(N_1)) - \sup_{t\in[0,1]}\left[H(t) + \frac{1}{\beta}\log\mathbb{E}\left(2^{-q(N_1-2\sqrt{t\beta}Z)}\right)\right]$$

$$= 1 - \sup_{t\in[0,1]}\left[H(t) + \frac{1}{\beta}\left(\mathbb{E}(q(N_1)) + \log\mathbb{E}\left(2^{-q(N_1-2\sqrt{t\beta}Z)}\right)\right)\right] \quad \text{(B1)} \blacksquare$$

**Special Case of Gaussian Noise**: Let $f(x) = \frac{1}{\sqrt{2\pi}\sigma} e^{\frac{-x^2}{2\sigma^2}}$ be the pdf of a centered Gaussian random variable of variance $\sigma^2$ and set $q(x) = \frac{\gamma \log e}{2} \left(\frac{x}{\sigma}\right)^2$. In this case clearly we have $\mathbb{E}(q(N_1)) = \frac{\log e}{2}\gamma$. To compute $\mathbb{E}\left(2^{-q(N_1 - 2\sqrt{t\beta}Z)}\right)$ in this case we need to do some more calculations for the Gaussian random variable:

If $Z$ is a standard Gaussian random variable and $\alpha, \beta$ are arbitrary real numbers, we have

$$\mathbb{E}\left(e^{-\frac{1}{2}\alpha Z^2}\right) = \frac{1}{\sqrt{\alpha+1}}$$

Thus, we have

$$\log \mathbb{E}\left(e^{-q(N_1 - 2\sqrt{t\beta}Z)}\right) = \log \mathbb{E}\left(e^{-q\left(\sqrt{\sigma^2 + 4t\beta}Z\right)}\right) = \log \mathbb{E}\left(e^{-\frac{\gamma}{2}\left(\frac{\sqrt{\sigma^2+4t\beta}}{\sigma}Z\right)^2}\right)$$

$$= \log\left(\frac{1}{\sqrt{1 + \frac{\gamma}{\sigma^2}(\sigma^2 + 4t\beta)}}\right) = -\frac{1}{2}\log\left(1 + \frac{\gamma}{\sigma^2}(\sigma^2 + 4t\beta)\right)$$

Hence the right-hand side of (B1) in this case becomes

$$1 - \sup_{t \in [0,1]} \left(H(t) + \frac{1}{2\beta}\left(\gamma \log e - \log\left(1 + \frac{\gamma}{\sigma^2}(\sigma^2 + 4t\beta)\right)\right)\right)$$

Thus, we get (9). ∎

**Appendix C**

**Tanaka's Bound Approaches Our Asymptotic Upper Bound as $\beta$ Increases**

In this appendix, we show that for large $\beta$ and the noise variance $\sigma^2 > 0$, Tanaka's formula is close to our bound $\frac{1}{2\beta} \log\left(1 + \frac{\beta}{\sigma^2}\right)$. Tanaka's formula can be rewritten as [12]:

$$C_m = \frac{1}{2\beta}\log\left(1 + \frac{\beta(1-m)}{\sigma^2}\right) + g(\lambda, m)\log(e)$$

In which,

$$g(\lambda, m) = \frac{\lambda}{2}(1+m) - \int \ln(\cosh(\sqrt{\lambda}\, Z + \lambda))D_Z$$

where $D_Z$ is the standard normal measure and

$$\lambda = \frac{1}{\sigma^2 + \beta(1-m)}$$

$$m = \int \tanh(\sqrt{\lambda}Z + \lambda)\, D_Z$$

Notice that $\lambda \leq \frac{1}{\sigma^2}$ and hence

$$m \leq \int_{-\sqrt{\lambda}}^{+\infty} \tanh(\sqrt{\lambda}Z + \lambda)D_Z \leq \int_{-\sqrt{\lambda}}^{+\infty} 1.D_Z \leq \int_{-\frac{1}{\sigma}}^{+\infty} 1.D_Z = Q\left(-\frac{1}{\sigma}\right) < 1$$

And therefore $\lim_{\beta \to \infty} \lambda = 0$.

for $x \geq 0$, $\tanh(x) \leq x$, thus we can rewrite the above equation as:

$$m < \int_{-\sqrt{\lambda}}^{+\infty} \tanh(\sqrt{\lambda}Z + \lambda)D_Z < \int_{-\sqrt{\lambda}}^{+\infty} (\sqrt{\lambda}Z + \lambda)D_Z < \frac{\sqrt{\lambda}}{\sqrt{2\pi}}e^{\frac{-\lambda}{2}} + \lambda < \frac{\sqrt{\lambda}}{\sqrt{2\pi}} + \lambda = O(\sqrt{\lambda}) \to 0$$

Using the Taylor expansion about 0, it can be easily seen that $\ln(\cosh x) = \frac{x^2}{2} + O(x^4)$, thus

$$g(\lambda, m) = \frac{\lambda}{2}(1+m) - \int \frac{(\sqrt{\lambda}Z + \lambda)^2}{2}D_Z + O(\lambda^2) = \frac{\lambda}{2} + O(\lambda\sqrt{\lambda}) - \frac{\lambda}{2} - \frac{\lambda^2}{2} + O(\lambda^2) = O(\lambda\sqrt{\lambda})$$

And since $\lambda \sim \frac{1}{\beta}$, consequently $\frac{g(\lambda, m)}{\frac{1}{2\beta}\log\left(1+\frac{\beta}{\sigma^2}\right)} \to 0$ and hence

$$C_m \sim \frac{1}{2\beta}\log\left(1 + \frac{\beta}{\sigma^2}\right) \blacksquare$$

**Appendix D**

Donsker-Varadhan inequality [25]: If $\mu$ and $\upsilon$ are two probability measures on a given space $\mathcal{X}$, then

$$\inf_{V:\mathcal{X} \to \mathbb{R}}\left[\ln \int e^V d\mu - \int V d\upsilon\right] = -\int \ln f\, d\upsilon$$

where $f = \frac{dv}{d\mu}$ is the Radon-Nikodym derivative of $v$ with respect to $\mu$. Infimum is attained at $V = \ln f$.

We proved that for any arbitrary function $q$,

$$\lim_{\substack{n/m=\beta \\ n,m\to\infty}} c(m,n,f) \geq 1 - \sup_{t\in[0,1]} \left[H(t) + \frac{1}{\beta}\left(\mathbb{E}(q(N_1)) + \log \mathbb{E}\left(2^{-q(N_1-2\sqrt{t\beta}Z)}\right)\right)\right] \quad (B1)$$

And hence

$$\lim_{\substack{n/m=\beta \\ n,m\to\infty}} c(m,n,f) \geq 1 - \inf_{q} \sup_{t\in[0,1]} \left[H(t) + \frac{1}{\beta}\left(\mathbb{E}(q(N_1)) + \log \mathbb{E}\left(2^{-q(N_1-2\sqrt{t\beta}Z)}\right)\right)\right]$$

But

$$\inf_{q} \sup_{t\in[0,1]} \left[H(t) + \frac{1}{\beta}\left(\mathbb{E}(q(N_1)) + \log \mathbb{E}\left(2^{-q(N_1-2\sqrt{t\beta}Z)}\right)\right)\right]$$

$$\geq \sup_{t\in[0,1]} \inf_{q} \left[H(t) + \frac{1}{\beta}\left(\mathbb{E}(q(N_1)) + \log \mathbb{E}\left(2^{-q(N_1-2\sqrt{t\beta}Z)}\right)\right)\right]$$

$$= \sup_{t\in[0,1]} \left[H(t) + \frac{1}{\beta}\inf_{q}\left[\left(\mathbb{E}(q(N_1)) + \log \mathbb{E}\left(2^{-q(N_1-2\sqrt{t\beta}Z)}\right)\right)\right]\right]$$

Now let $\mu$ and $v$ be probability measures induced on $\mathbb{R}$ from the random variables $N_1 - 2\sqrt{t\beta}Z$ and $N_1$ respectively, and let $V = -\frac{1}{\log e}q$. Then Donsker-Varadhan inequality implies that

$$\inf_{q}\left[\left(\mathbb{E}(q(N_1)) + \log \mathbb{E}\left(2^{-q(N_1-2\sqrt{t\beta}Z)}\right)\right)\right] = (-\log e)\int \ln f \, dv = (-\log e)\mathbb{E}(\ln f(N_1))$$

where $= \frac{dv}{d\mu}$. For the Gaussian noise, $f(x) = \sqrt{1 + \frac{4t\beta}{\sigma^2}}\, e^{-\frac{1}{2\sigma^2(\sigma^2+4t\beta)}\cdot 4t\beta \cdot x^2}$, and thus

$$\mathbb{E}(\ln f(N_1)) = \mathbb{E}\left(\frac{1}{2}\ln\left(1+\frac{4t\beta}{\sigma^2}\right) - \frac{1}{2}\frac{4t\beta}{\sigma^2(\sigma^2+4t\beta)}N_1^2\right) = \frac{1}{2}\left(\ln\left(1+\frac{4t\beta}{\sigma^2}\right) - \frac{4t\beta}{(\sigma^2+4t\beta)}\right)$$

Therefore we have

$$\inf_{q} \sup_{t\in[0,1]} \left[H(t) + \frac{1}{\beta}\left(\mathbb{E}(q(N_1)) + \log \mathbb{E}\left(2^{-q(N_1-2\sqrt{t\beta}Z)}\right)\right)\right]$$

$$\geq \sup_{t\in[0,1]} \left[H(t) + \frac{\log e}{2\beta}\left(\frac{4t\beta}{(\sigma^2+4t\beta)} - \ln\left(1+\frac{4t\beta}{\sigma^2}\right)\right)\right]$$

Hence the best lower bound obtainable from (B1) is not better than

$$1 - \sup_{t \in [0,1]} \left[ H(t) + \frac{\log e}{2\beta} \left( \frac{4t\beta}{(\sigma^2 + 4t\beta)} - \ln\left(1 + \frac{4t\beta}{\sigma^2}\right) \right) \right] \quad (D1) \blacksquare$$

**Acknowledgement**

We would like to sincerely thank A. Amini, G.H. Mohimani, B. Akhbari, and M.H. Yasaee for their helpful comments. We are specifically indebted to the two anonymous reviewers and the associate editor for their helpful comments; in fact, Dr. Gerhard Kramer has brought to our attentions the similarity of our upper bound for the noiseless case with that of T- codes developed by Chang and Weldon [24], and the proof that our conjectured upper bound is a true upper bound for the noiseless case (Section 4) based on the works of Shepp and Olgin [27] and the two corresponding papers [28]-[29].

## Biographies

**Dr. Kasra Alishahi** received the B.S., M.S., and Ph.D. degrees all from the Department of Mathematical Sciences, Sharif University of Technology, Tehran, Iran, in 2000, 2002, and 2008, respectively. He is currently an Assistant Professor with Sharif University of Technology. His research interests are stochastic processes and stochastic geometry. Dr. Alishahi was a recipient of a gold medal in the International Mathematical Olympiad in 1998.

**Dr. Farokh Marvasti** (S'72–M'74–SM'83) received the B.S., M.S., and Ph.D. degrees all from Renesselaer Polytechnic Institute, Troy, NY, in 1970, 1971, and 1973, respectively. He has worked, consulted, and taught in various industries and academic institutions since 1972. Among which are Bell Labs, University of California Davis, Illinois Institute of Technology, University of London, King's College. He is currently a professor with Sharif University of Technology, Tehran, Iran, and the Director of the Advanced Communications Research Institute (ACRI). He has approximately 60 journal publications and has written several reference books. His latest book is *Nonuniform Sampling: Theory and Practice* (New York: Kluwer, 2001). Dr. Marvasti was one of the Editors and Associate Editors of the IEEE TRANSACTIONS ON COMMUNICATIONS AND SIGNAL PROCESSING from 1990 to 1997. He is also a Guest Editor of the Special Issue on Nonuniform Sampling for the *Sampling Theory and Signal and Image Processing Journal*

**Mr. Vahid Aref** (S'07) was born in 1984. He received the B.S. and M.S. degrees all from the Department of Electrical Engineering, Sharif University of Technology, Tehran, Iran, in 2006 and 2008, respectively. He is planning to continue his PhD at EPFL in Lausanne, Switzerland in September of 2009. Mr. Aref received a silver medal in the National Physics Olympiad competition in 2001.

**Mr. Pedram Pad** (S'03) was born in Iran in 1986. He is a double major B.S. degree student of electrical engineering and pure mathematics at Sharif University of Technology, Tehran, Iran. He is also a member of Advanced Communications Research Institute (ACRI). Mr. Pad received a gold medal in the National Mathematical Olympiad competition in 2003.